\documentclass[floatfix,aps,prl,twocolumn,superscriptaddress,showpacs,nofootinbib,reprint]{revtex4-1}
\pdfoutput=1
\usepackage{float}
\usepackage{color}
\usepackage{graphicx}
\usepackage{ulem}
\usepackage{epstopdf}
\usepackage{amsfonts}
\usepackage{amsmath}
\usepackage{mathtools}
\usepackage{bbold}
\usepackage[toc,page]{appendix}
\usepackage{url}

\newcommand{\omn}{\omega_{\text{\tiny{NN}}}}
\newcommand{\omb}{\omega_{\text{\tiny{2B}}}}
\newcommand{\omemt}{\omega_{\text{\tiny{EMT}}}}
\newcommand{\perc}{\mathcal{P_{\omega_{\text{\tiny{NN}}}}}}

\newcommand{\percemt}{\mathcal{P_{\omega_{\text{\tiny{EMT}}}}}}
\newcommand{\x}{\mathbf{x}}

\newcommand\chout{\bgroup\markoverwith{\textcolor{red}{\rule[0.5ex]{2pt}{1.0pt}}}\ULon}
\newcommand\fedout{\bgroup\markoverwith{\textcolor{blue}{\rule[0.5ex]{2pt}{1.0pt}}}\ULon}
\newcommand\grzout{\bgroup\markoverwith{\textcolor{green}{\rule[0.5ex]{2pt}{1.0pt}}}\ULon}

 \usepackage{pdfpages}
\usepackage{pgffor}
\makeatletter
\AtBeginDocument{\let\LS@rot\@undefined}
\makeatother
\begin{document}

\title{Non-equilibrium dynamics of isostatic spring networks}

\author{Federico S. Gnesotto}
    \thanks{These authors contributed equally}
\affiliation{Arnold-Sommerfeld-Center for Theoretical Physics and Center for
  NanoScience, Ludwig-Maximilians-Universit\"at M\"unchen,
   D-80333 M\"unchen, Germany.}
\author{Benedikt M. Remlein}
    \thanks{These authors contributed equally}
\affiliation{Arnold-Sommerfeld-Center for Theoretical Physics and Center for
  NanoScience, Ludwig-Maximilians-Universit\"at M\"unchen,
   D-80333 M\"unchen, Germany.}
\author{Chase P. Broedersz}
\email{C.broedersz@lmu.de}
\affiliation{Arnold-Sommerfeld-Center for Theoretical Physics and Center for
  NanoScience, Ludwig-Maximilians-Universit\"at M\"unchen,
   D-80333 M\"unchen, Germany.}

\date{\today}

\begin{abstract}
Marginally stable systems exhibit rich critical mechanical behavior. Such isostatic assemblies can be driven out of equilibrium by internal activity,  but it remains unclear how the isostatic and critical nature of such systems affects their non-equilibrium dynamics. Here, we investigate the influence of the isostatic threshold on the non-equilibrium dynamics of active diluted spring networks. In our model, heterogeneously distributed active noise sources drive the system into a non-equilibrium steady state. We quantify the non-equilibrium dynamics of nearest-neighbor network nodes by the characteristic cycling frequency $\omega$---a measure of the circulation of the associated phase space currents. The distribution of these nearest-neighbor cycling frequencies exhibits critical scaling, which we describe using a mean-field approach.  Overall, our work provides a theoretical approach to elucidate the role of marginality in active disordered systems. 

\end{abstract}
\maketitle
\noindent 
Isostaticity has been central in providing a unified understanding of the mechanics of soft disordered systems~\cite{Liu1998,vanHecke2010,Lubensky2015,Broedersz2014Rev}. A system is isostatic when its degrees of freedom are exactly balanced by its internal constraints, poising the system at the verge of mechanical stability~\cite{Maxwell,Jacobs1996}. Examples of such marginal matter include colloidal suspensions, granular packings and foams near the jamming transition~\cite{Cates1998,Liu2010}, as well as spring lattices~\cite{Thorpe83,Wyart2008,Mao2010} and biological  fiber networks~\cite{Heussinger97,Broedersz2011,Broedersz2014Rev}. Such systems display critical behavior near the isostatic connectivity point such as universality, a nonlinear elastic response, and diverging strain fluctuations \cite{Broedersz2011,Sheinman2012,Broedersz2014Rev,Lubensky2015,Sharma2016}. Recently, a variety of non-equilibrium dynamics has been reported in active soft matter, ranging from  biopolymer assemblies with molecular motors~\cite{Alvarado2013,Brangwynne2008,Schaller2010,Huber2018} and vibrated polar disks~\cite{Deseigne2010} to living systems such as cells, tissues and bacterial populations~\cite{Angelini2011,Bi2015,Thutupalli2015,Delarue2016}.  Although there is evidence that such collective dynamics occurs in the vicinity of an isostatic threshold~\cite{Alvarado2013,Bi2015,Tan2018,Alvarado2017}, it remains unclear to what extent the critical nature of these systems affects their non-equilibrium  dynamics~\cite{Woodhouse2018Arx,Wan2018}. More generally, a theoretical framework for the non-equilibrium stochastic dynamics of marginally stable systems is still lacking.

\begin{figure}[h!]
\centering
 \includegraphics[width=0.9 \columnwidth]{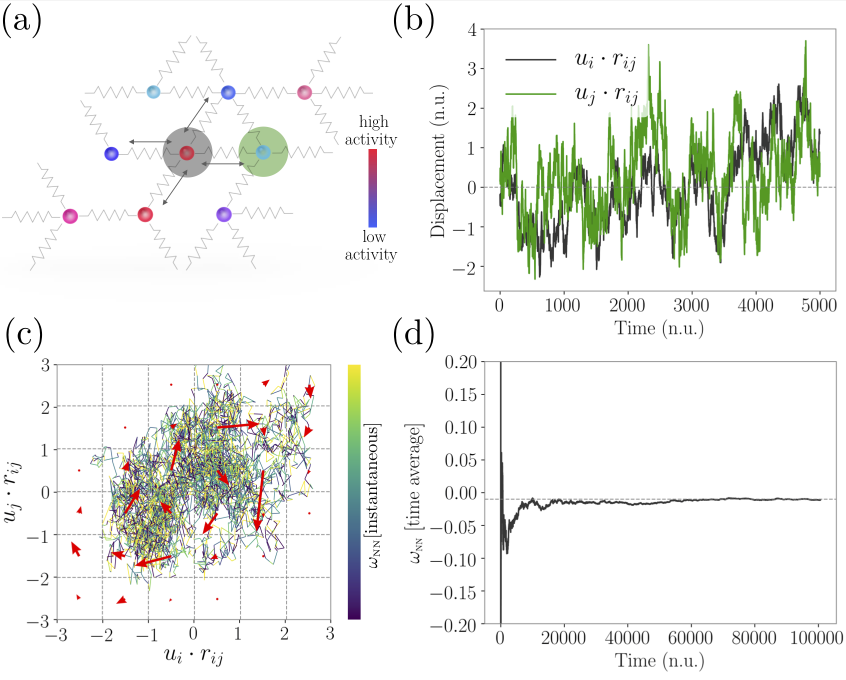}
  \caption{(a) Schematic of a driven disordered network of beads and springs: bead colors  indicate local active noise amplitudes, increasing from blue to red. We consider the coupled stochastic dynamics of nearest-neighbor beads (double arrows): for all such pairs we calculate the cycling frequency $\omn$. (b) The  displacements of two nearest-neighbor beads projected along the associated bond directions (panel (a)) as a function of time. All quantities are plotted in natural units (see main text).  (c) Phase space trajectory (colored by the instantaneous cycling frequency) and probability current (red arrows) for the data in panel (b). (d) The time-averaged cycling frequency converges to a non-zero value as $t \rightarrow \infty$.}
    \label{Fig1}
\end{figure}

Here we propose a simple model for an internally driven marginal elastic system:  an active randomly diluted triangular network with tunable connectivity. This model allows us to systematically investigate the role that the  isostatic connectivity point plays in controlling the system's non-equilibrium fluctuations. In our model we consider a heterogeneous distribution of active sources at the nodes: activity differences drive the network out of equilibrium, thereby breaking detailed balance (Fig.~\ref{Fig1}(a)). At steady state, broken detailed balance is associated with the presence of circulating probability currents in the phase space of configurational observables~\cite{Battle2016,Gnesotto2018}. Consequently, the phase space trajectory of a pair of degrees of freedom circulates on average  with a cycling frequency $\omega$ (Fig. \ref{Fig1}(c)-(d)). We employ this measure to quantify broken detailed balance between pairs of directly connected nodes~\cite{Weiss2003,Gladrow2016,Gladrow2017}. Interestingly, as we lower the  network connectivity, we find that the distribution of cycling frequencies changes drastically  near the isostatic threshold. To characterize this transition, we use the $68_{\rm th}$ percentile of this distribution as an order parameter. We support our choice by demonstrating that the $68_{\rm th}$ percentile obeys characteristic scaling laws in the form of a homogeneity relation, and we develope a mean-field theory for this scaling behavior. Taken together, our results demonstrate how isostaticity can control the non-equilibrium dynamics in disordered systems. 

We start our analysis by considering an elastic network of beads connected by springs arranged on a triangular lattice in 2D. The lattice is immersed in an incompressible newtonian fluid at temperature $T$ and all springs have elastic constant $k$ and rest length $\ell_0$ (Fig. \ref{Fig1}(a)). To tune network connectivity, we randomly dilute bonds with probability $1-p$, with $p \in [0,1]$, resulting in a network of average connectivity $z=6p$. The overdamped stochastic equation for the position $\mathbf{x}_i$ of each node reads 
\begin{equation}
\gamma \frac{d \mathbf{x}_i}{dt}(t)=-\sum_{\langle i,j \rangle}k_{i,j}(\lVert \mathbf{x}_{i,j}(t) \rVert-\ell_0)\hat{\mathbf{x}}_{i,j} (t) + a_i \boldsymbol{\mathbf{\eta}}_i(t) \, ,
\label{eq:EOMfull}
\end{equation}
where $\gamma$ is the drag coefficient of each bead in the fluid, $k_{i,j}=k$  if the bond is present or $k_{i,j}=0$ if the bond is removed, $\mathbf{x}_{i,j}=\mathbf{x}_{i}-\mathbf{x}_{j}$, and $\hat{\mathbf{x}}_{i,j}$ is the corresponding unit vector. Note, we neglect hydrodynamic interactions between beads and use fixed boundary conditions to prevent rigid body translation and rotation.

Internal fluctuating forces acting on the beads are described as Gaussian white noise. Thus, for nodes $i$ and $j$, $\langle \eta_{ix} (t) \eta_{jy}(t') \rangle=\delta_{ij} \, \delta_{xy} \, \delta(t-t')$ and $\langle {\eta}_{ix} \rangle =\langle \eta_{iy} \rangle =0$. Here, the white noise amplitude includes both thermal and active fluctuations: $a_i=\sqrt{2 \gamma k_B (T+|\alpha_i|})$. Importantly, while the amplitude of the thermal contribution is homogeneous throughout the system, the amplitude of active noise may be heterogeneous and will be described using quenched disorder. Specifically, we draw the amplitudes $\alpha_i$ from a normal distribution with average $\mu_{\alpha}$ and variance $\sigma_{\alpha}^2$, such that $\sigma_{\alpha} \ll T+\mu_{\alpha} $. This approach allows us to investigate a general scenario of marginally stable system driven out of equilibrium by internal driving. Note, when $\sigma_{\alpha}=0$, the system obeys equilibrium dynamics. In what follows we use natural units,  measuring time in units of $\gamma/k$, lengths in units of $\ell_0$ and temperature (and activity) in units of $k \ell_0^2/k_B$, leaving only $a_i=\sqrt{2(T+|\alpha_i|)}$ free in Eq.~\eqref{eq:EOMfull}.

\begin{figure}[t!]
\centering
  \includegraphics[width=0.9 \columnwidth]{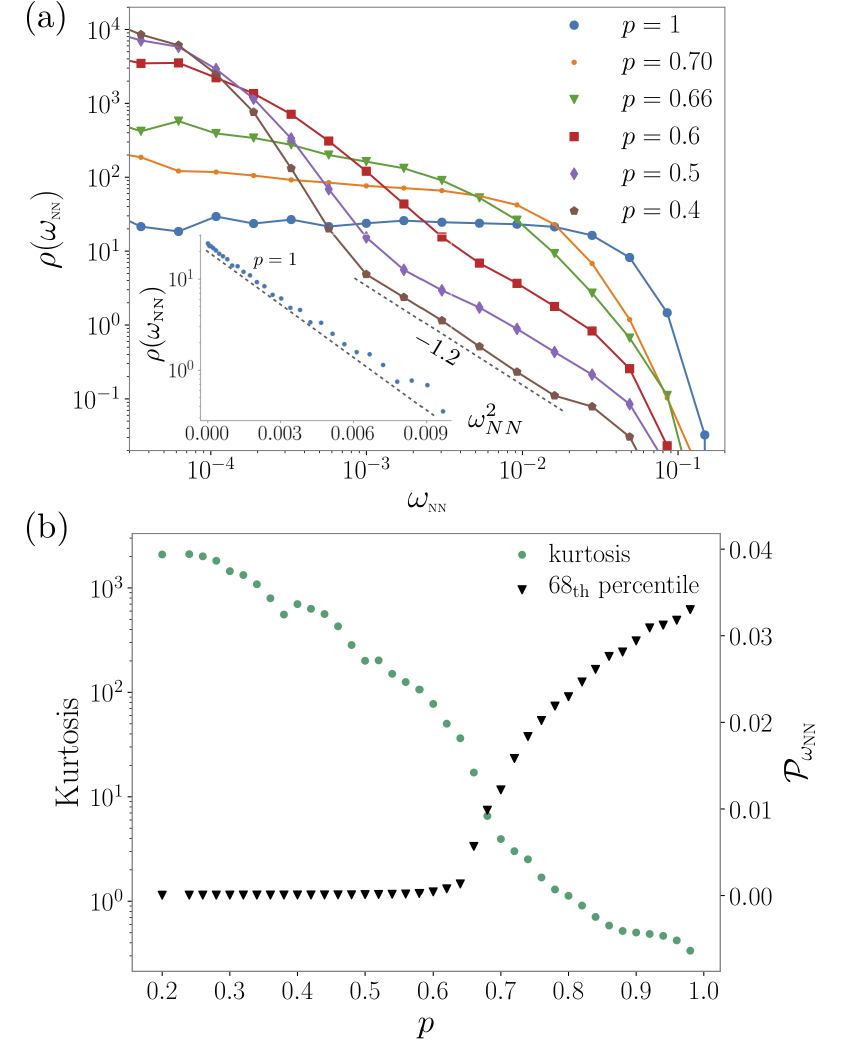}

  \caption{(a) The probability density of cycling frequencies $\rho(\omn)$, for various $p$. Inset shows that  the distribution is approximately Gaussian when the network is barely diluted (i.e. $p \lesssim 1$). (b) The kurtosis (orange triangles) and the $68_{\rm th}$ percentile $\perc$ (green circles) of the cycling frequency distribution, for different values of $p$. These results are obtained by simulating a $100\times100$ system with $\sigma_{\alpha}=10^{-5}, \, T+\mu_{\alpha}=10^{-4}$.}
    \label{Fig2}
\end{figure}
To simulate our model we employ a  Brownian Dynamics approach: this allows us, for instance, to track the displacements of two neighboring beads in the network. Two typical trajectories are shown in Fig.~\ref{Fig1}(b). Although we could, in principle, compute the  probability current field (red arrows in Fig.~\ref{Fig1}(c)), a simpler but still meaningful way of quantifying the non-equilibrium dynamics of these two beads is  the cycling frequency - the average number of revolutions of the trajectory  around the origin in phase space per unit time. In general, the instantaneous cycling frequency is a random variable (Fig. \ref{Fig1}(c)), but its time-averaged value, $\omn$, assumes a well defined value in the long time limit (Fig.~\ref{Fig1}(d))~\cite{Weiss2003,Gladrow2016,Gladrow2017}, which is related to the entropy production rate~\cite{Mura2018}. Thus, by determining the cycling frequencies, we assign a simple  pseudoscalar measure of non-equilibrium to each pair of connected neighboring beads in our network. The cycling frequencies for distinct bead pairs will in general not only differ because of the heterogeneous activities, but also because of the disordered network structure. 

To investigate the interplay between internal driving and disorder, we determine the  (symmetric) probability distribution, $\rho(\omega_{\rm NN})$, of nearest-neighbor cycling frequencies $\omega_{\rm NN}$ for different values of the dilution parameter $p$ (Fig.~\ref{Fig2}(a)). Given that the activity magnitudes are drawn from a normal distribution, it is perhaps not surprising that at $p=1$ the cycling frequencies are also normally distributed (inset Fig.~\ref{Fig2}(a)). Interestingly however, as we dilute the network down to the isostatic threshold for this system size ($p_c \approx 0.65$~\cite{Suppl}), a pronounced peak develops around the origin. Furthermore, slowly decaying heavy tails with an apparent power-law dependence characterize the distribution for large $\omega_{\rm NN}$ (Fig. \ref{Fig2}(a)); correspondingly, the kurtosis of the distribution increases markedly as we lower $p$ (Fig. \ref{Fig2}(b)). These results indicate that the standard deviation is not appropriate to characterize the   cycling frequency distribution. Instead, we use the 68$_{\rm th}$ percentile,  $\perc$, of the distribution; $\perc$ is non-zero in the rigid (ordered) phase $p>p_c$, while it falls to zero continuously when $p < p_c$.

\begin{figure}[t]
\centering
  \includegraphics[width=0.9 \columnwidth]{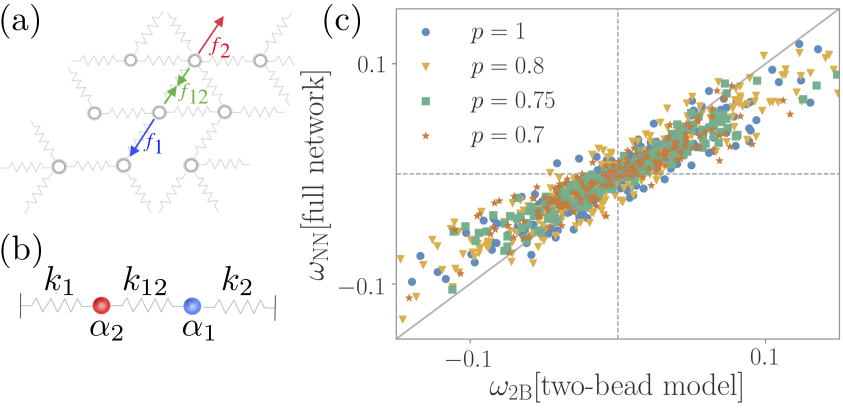}
  \caption{ (a) Two monopole forces $f_1$ and $f_2$ and a dipole $f_{12}$ are applied at two neighboring network nodes. The response to the forces yields three different effective spring constants $k_1, k_2, k_{12}$. (b) Two-bead model with activities equal to those of the nodes in the full network. (c) Scatter plot of the cycling frequencies $\omn$ calculated for each pair of neighboring nodes for full networks with varying $p$ (y-axes) and the corresponding estimate from the two-bead model $\omega_{\text{\tiny{2B}}}$  (x-axes).}   
   \label{Fig3}
\end{figure}

To provide intuition for what features determine the distribution of cycling frequencies, we first ask what sets the local value of $\omega_{\rm NN}$. It is instructive to start by considering a purely elastic network where all thermal and active fluctuations have been suppressed. Specifically, we determine the linear response of the system to  three configurations of forces applied to a pair of neighboring nodes in the network: two monopoles and one dipole force, all directed along the lattice bond connecting the two beads, as in Fig. \ref{Fig3}(a). By assessing the directed displacements of the beads for each of the three force configurations, we measure three different elastic responses. Using these  responses, we map the local mechanical response in the disordered network onto an effective one-dimensional two-bead model (Fig. \ref{Fig3}(b)) with spring constants $k_1, k_2, k_{12}$. These spring constants are set such that the effective system retains the same elastic response to the three force configurations as the local response in the full network.

While this procedure works well for a purely mechanical network, the mapping is in general not valid for the stochastic dynamics of the active system~\cite{Mehl2012,Uhl2018}. However, we can, as a first approximation, neglect the active fluctuations of all other beads in the network and insert in our two-bead model only the activities of the considered pair of nodes (Fig.~\ref{Fig3}(b)).  
By ascribing the same activities of the pair of nodes in the network to the nodes of the effective two-bead model, we can make a prediction for the cycling frequencies. In the limit of small activity difference, i.e. $\alpha_1=T+\alpha+\delta \alpha, \, \alpha_2=T+\alpha$ with $\delta \alpha \ll T+\alpha$, the cycling frequency for the two-bead model reads~\cite{Suppl}
\begin{equation}
\omega_{\text{\tiny{2B}}} \approx k_{12}\frac{\sqrt{k_{12}(k_1+k_2)+k_1k_2}}{k_1+2k_{12}+k_2}\frac{\delta \alpha}{T+\alpha}.\label{eq:Two_Bead_3k}
\end{equation}
The cycling frequencies $\omn$ for every pair of neighbors in the full disordered network agree well on a case by case basis with the estimates of the two-bead model  $\omb$ (Fig.~\ref{Fig3}(c)). However, the cycling frequencies $\omb$ predicted by the two-bead model are, in absolute value, larger than the ones obtained from our network simulations. We attribute this effect to the activities of the other network nodes~\cite{Suppl}, which are excluded in this simple model. Nonetheless, these results indicate that the local mechanical response together with the local activity difference set the scale of the local cycling frequency. 

\begin{figure}[t]
\centering
  \includegraphics[width=0.9 \columnwidth]{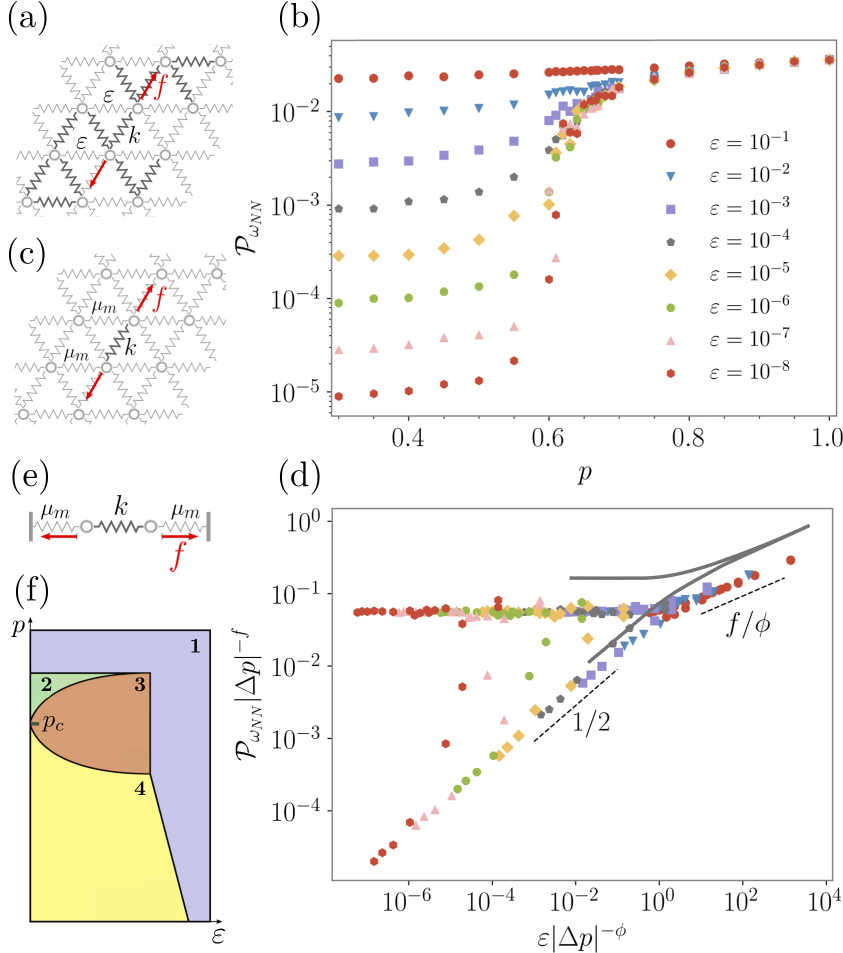}
  \caption{(a) Schematic of network with $\varepsilon$-bond replacement. (b) The $68_{\rm th}$ percentile, $\perc$ for different $p$ and $\varepsilon$. These results were obtained by solving Eq.~\eqref{eq:Lyapunov} for a triangular lattice of size $W=40\times40$ and  $\sigma_{\alpha}=10^{-4}, \, T+\mu_{\alpha}=10^{-3}$.(c) Schematic of effective medium, with all bonds replaced by effective medium springs except for the bonds where the dipole is applied. (d) Scaling of the $68_{\rm th}$ percentile $\mathcal{P_{\omn}}(p,\varepsilon)=|\Delta p|^f P(\varepsilon|\Delta p|^{-\phi})$  around the critical point. The solid line indicates the mean-field prediction. (e) Effective two-bead model with external spring constants $\mu_m$. (f) Schematic phase diagram: (1)~$\percemt \sim k$; (2)~$\percemt \sim k^{1/2}|\Delta p|^{1/2}$; (3)~$\percemt \sim k^{3/4}\varepsilon^{1/4}$; (4)~$\percemt \sim k^{1/2}\varepsilon^{1/2}|\Delta p|^{-1/2}$.}
    \label{Fig4}
\end{figure}

Studying analytically the full system described by Eq.~\eqref{eq:EOMfull} is arduous, as nonlinearities may become increasingly more important when the system is diluted and driven out of equilibrium by large noise. However, in the limit of modest driving, the elastic contribution to the force in our model can be linearized, which results in a simplified equation of motion (in natural units)
\begin{equation}
\frac{d \mathbf{u}_i}{dt}(t)=-\sum_{\langle j \rangle} A_{ij}\mathbf{u}_{j} +a_i  \boldsymbol{\mathbf{\eta}}_i(t) \, ,
\label{eq:LinearizedEOM}
\end{equation}
with $\mathbf{u}_{i}$ representing the  displacement of node $i$ from its rest position and the elastic-matrix $A$ being defined as
\begin{equation}
A_{i\alpha j\beta} = \begin{cases}
-k_{i,j} \hat{r}_{i,j\alpha}\hat{r}_{i,j\beta},&i \neq j\\
\sum\limits_{n\neq i}k_{in}\hat{r}_{i,n\alpha}\hat{r}_{i,n\beta},&i = j
\end{cases}\, ,
\label{eq:AMatrix}
\end{equation}
where $\hat{\mathbf r}_{i,j}$ is the unit vector connecting the rest positions of nodes $i$ and $j$ and  greek indices denote  cartesian components.
For such a linear system, the steady-state covariance matrix $C_{i\alpha j\beta}=\langle {u}_{i\alpha} {u}_{j\beta} \rangle$ satisfies the Lyapunov equation~\cite{LyapunovGeneral}:
\begin{equation}
A C + CA^T = - 2 D \, ,
\label{eq:Lyapunov}
\end{equation}
with $D$ being the diffusion matrix with elements $D_{i\alpha j\beta}=\frac{1}{2}a_i^2\delta_{ij}\delta_{\alpha\beta}$.
While for a fully connected network $A$ is invertible,  zero-energy modes with diverging relaxation time start to appear in the system as we remove bonds from the network and approach the isostatic point $p_c$. Infinite relaxation times lead to divergences in the elements of the covariance matrix. To avoid these divergences, it is convenient to insert a weak $\varepsilon$-spring of elastic constant $\varepsilon \ll 1$ ($\varepsilon$ is in units of $k$) whenever a $k$-spring is removed, as sketched in Fig. \ref{Fig4}(a) ~\cite{Garboczi1986,Wyart2008}. In the limit $\varepsilon \to 0$ we expect to recover the dynamics of the simulated network. This dilute-and-replace procedure allows us to stabilize the zero modes in a controlled way, thereby avoiding singularities of the covariance matrix. Because of the 6-fold rotational symmetry of the lattice, we obtain the cycling frequencies by considering a specific direction, namely for $x$-displacements of nodes connected by $x$-directed bonds~\cite{Mura2018}:
\begin{equation}
\omega_{ij} = \frac 12 \frac{\left(AC - CA^T\right)_{i x j x}}{\sqrt{C_{i x i x}C_{j x j x} - C_{i x j x}C_{j x i x}}} \, .
\label{eq:Omegas}
\end{equation} 
In the limit of modest activity: $\sigma_{\alpha} \ll (T+\mu_\alpha )\ll 1$, the full non-linear model (Eq.~\eqref{eq:EOMfull}) is well approximated by its linearized version (Eq.~\eqref{eq:LinearizedEOM}). As a result, we can numerically obtain the distributions of Fig.~\ref{Fig2}(a) and in particular the observable $\perc$. Moreover, the introduction of a soft spring constant $\varepsilon$ allows us to stabilize the rigidity of our network also below $p_c$. Using this approach, we find that the characteristic change in the shape of $\rho(\omn)$ near the critical point $p_c$  (see Fig.~\ref{Fig2}(b)), is reflected by a sharp but continuous decrease of $\perc$, as shown in  Fig.~\ref{Fig4}(b) for varying  $\varepsilon$. 

Because the network is stabilized by the soft $\varepsilon$-springs, the jump in $\perc$ becomes less pronounced for larger values of $\varepsilon$. Thus, it appears as if $\varepsilon$ acts as a scaling field that takes the system away from criticality~\cite{Wyart2008,Broedersz2011}.  To test this idea, we investigate if $\perc$ obeys a homogeneity relation of the  form
\begin{equation}
 \mathcal{P_{\omn}}(p,\varepsilon)=|\Delta p|^f P_{\pm}(\varepsilon|\Delta p|^{-\phi}) \, ,
 \label{eq:Scaling_Perc}
\end{equation}
where $\Delta p=p-p_c$ and $P_{\pm}$ is a universal function. To this end,  we rescale the data for different $\varepsilon$ and $p$ according to this relation and observe a good collapse, as shown in Fig.~\ref{Fig4}(d). Based on this analysis, we identify three distinct scaling regimes: a super-critical, $k$-dominated regime where  $\perc \sim |\Delta p|^f$, a critical regime $\perc \sim \varepsilon^{f / \phi}$, and a sub-critical one where $\perc \sim \varepsilon^{1/2}|\Delta p|^{f-\phi/2}$. Empirically, we observe a reasonable collapse with the exponents $f=0.45 \pm 0.05, \, \phi=1.8 \pm 0.2$.
\smallskip

To provide insight into the origin of the critical scaling of the cycling frequency distribution, we build on the two-bead model (Fig.~\ref{Fig3}) to develop a mean-field approach. In particular, we use an effective medium theory (EMT)  to predict the statistical properties of the cycling frequencies in our system. Importantly, we anticipate this approach to work  well in the low activity limit because of the structure of Eq.~\eqref{eq:Two_Bead_3k}: fluctuations in the elastic constants of the two-bead model only appear as a second order correction to the cycling frequency \cite{Suppl}. Thus, a mean-field prediction of the elastic constants $k_1$, $k_2$, and $k_{12}$ of the two bead model, should lead to an accurate estimate of the cycling frequencies for the full network. 

The idea underlying the EMT is to map a lattice with randomly diluted bonds onto a network with uniform bond-stiffness $\mu_m$~\cite{Feng1985}. This is accomplished by requiring equal elastic response between the effective medium and the disordered network when applying a dipole-force between two nodes, as illustrated in Fig.~\ref{Fig4}(a)-(c). This requirement leads to the following self-consistency equation
\begin{equation}
\left<\frac{\mu_m - k}{\mu_m/a^* - \mu_m +k}\right>= 0\, ,
\label{eq:Self_Consistency}
\end{equation}
where the average $\langle \cdot \rangle$ is taken over the distribution of stiffnesses of the network bonds. In our case the distribution of bond stiffnesses is binary $P(k_{i,j} = k) = p~,~~~P(k_{i,j} = \varepsilon) = 1-p~$, and the constant  $a^* = p_c =  2/3$ for a triangular network~\cite{Feng1985}.  By solving Eq.~\eqref{eq:Self_Consistency} for our system, we obtain the effective spring constant $\mu_m$~\cite{Garboczi1986}.

In the next step, we map the effective network onto a two-bead model, by requiring the dipole-response of the network shown in Fig.~\ref{Fig4}(c) to be equivalent to the response of the system of two beads shown in Fig.~\ref{Fig4}(e); this amounts to demanding that the two external springs  both have stiffnesses $\mu_m$.
Finally, we employ Eq.~\eqref{eq:Two_Bead_3k} in the small activity limit $\delta \alpha / (T+\alpha) \ll 1$ to obtain an analytical estimate of the cycling frequency~\cite{Suppl}
\begin{equation}
\omemt \approx \frac{k}{k+\mu_m}\sqrt{\mu_m(2k + \mu_m)} \cdot \frac{\delta \alpha}{2(T+\alpha)} \, .
\label{eq:TwoBead_eff_omega}
\end{equation}
If we choose $\alpha_1$ and $\alpha_2$---the activities of the two-bead model---to be distributed as the activities of the full network, we obtain the cycling frequency distribution $\rho(\omemt)$, from which we numerically calculate the $68_{\rm th}$ percentile $\percemt$. 
 This mean-field model successfully predicts the scaling of the $68_{\rm th}$ percentile for the original diluted network with exponents $f=1/2$, $\phi=2$  ~\cite{Suppl}, as demonstrated by the solid curve of Fig.~\ref{Fig4}(d). Note that the two-bead overestimate of the cycling frequencies already present in Fig.~\ref{Fig3}(c), is reflected here in the small but constant shift of the solid curve relative to the data in Fig.~\ref{Fig4}(d). The various phases and their boundaries predicted by this mean-field model are summarized in the phase diagram in Fig.~\ref{Fig4}(f) (see \cite{Suppl}). 

Our analytical approach captures the scaling of the order parameter $\percemt$ as well as the associated critical exponents. 
More than confirming our scaling ansatz Eq.~\eqref{eq:Scaling_Perc} (solid line Fig.~\ref{Fig4}(d)), this intuitive analytic approach provides insight into how  the non-equilibrium dynamics of a disordered marginal system can be understood by employing a mean-field model.

In conclusion, we have determined theoretically how the dynamics of actively driven elastic networks are governed by the vicinity to an isostatic critical point. This was accomplished by using simple experimentally accessible quantities such as the non-equilibrium cycling frequencies between pairs of nodes. These cycling frequencies are directly related to other non-equilibrium measures, such as the entropy  production rate~\cite{Mura2018}. Our results provide an important step towards establishing a general framework to probe the dynamics of disordered non-equilibrium systems to guide experimental studies of active marginal matter in biological~\cite{Alvarado2013,Bi2015,Tan2018,Wan2018} and synthetic systems~\cite{Narayan2007,Deseigne2010,Palacci2013}. 

\begin{acknowledgments}
We thank G. Gradziuk, K. Miermans, F. Mura, and P. Ronceray for helpful discussions.
This work was supported by the German Excellence Initiative via
 the program NanoSystems Initiative Munich (NIM) and   the Deutsche Forschungsgemeinschaft   (DFG)  Grant  GRK2062/1.
\end{acknowledgments}
\bibliography{Ref_BDB_Critical}


\section*{Supplementary material}

# Supplement: Non-equilibrium dynamics of isostatic spring networks

Federico S. Gnesotto,[1, *] Benedikt M. Remlein,[1, *] and Chase P. Broedersz[1, †]

[1]*Arnold-Sommerfeld-Center for Theoretical Physics and Center for NanoScience,
Ludwig-Maximilians-Universität München, D-80333 München, Germany.*

(Dated: September 12, 2018)

## DERIVATION OF EQ. (2)

In this supplement we derive the expression for the cycling frequency of the system depicted in Fig. 3(b) of the main text. The linear equation of motion for the displacements $u_i$ ($i = 1, 2$) of the two beads is

$$\frac{d}{dt}\begin{pmatrix} u_1 \\ u_2 \end{pmatrix} = A \begin{pmatrix} u_1 \\ u_2 \end{pmatrix} + \begin{pmatrix} a_1 \eta_1 \\ a_2 \eta_2 \end{pmatrix} \tag{S1}$$

where

$$A = \begin{pmatrix} -(k_1 + k_{12}) & k_{12} \\ k_{12} & -(k_{12} + k_2) \end{pmatrix}, \tag{S2}$$

$\langle \eta_i(t) \eta_j(t') \rangle = \delta_{ij} \delta(t - t')$ and $a_i = \sqrt{2\alpha_i}$ ($i \in \{1, 2\}$) is the amplitude of the Gaussian white noise. This choice leads to the diffusion matrix

$$D = \begin{pmatrix} \alpha_1 & 0 \\ 0 & \alpha_2 \end{pmatrix}, \tag{S3}$$

which can then be inserted in the Lyapunov-equation, $AC + CA^T = -2D$, to solve for the covariance matrix

$$C = \begin{pmatrix} \frac{2\left((\alpha_2 - \alpha_1)k_{12}^2 + (k_{12} + k_2)(k_1 + 2k_{12} + k_2)\alpha_1\right)}{(k_1 + 2k_{12} + k_2)(k_{12}k_2 + k_1(k_{12} + k_2))} & \frac{2k_{12}((k_{12} + k_2)\alpha_1 + (k_1 + k_{12})\alpha_2)}{(k_1 + 2k_{12} + k_2)(k_{12}k_2 + k_1(k_{12} + k_2))} \\ \frac{2k_{12}((k_{12} + k_2)\alpha_1 + (k_1 + k_{12})\alpha_2)}{(k_1 + 2k_{12} + k_2)(k_{12}k_2 + k_1(k_{12} + k_2))} & \frac{2\left((\alpha_1 + \alpha_2)k_{12}^2 + (3k_1 + k_2)\alpha_2 k_{12} + k_1(k_1 + k_2)\alpha_2\right)}{(k_1 + 2k_{12} + k_2)(k_{12}k_2 + k_1(k_{12} + k_2))} \end{pmatrix}. \tag{S4}$$

By using Eq. (6) from the main text for $i = 1$ and $j = 2$, we obtain the cycling frequency

$$\omega_{12} = (\alpha_1 - \alpha_2) k_{12} \sqrt{\frac{k_1(k_{12} + k_2) + k_{12}k_2}{4\alpha_1 \alpha_2 k_{12}(k_1 + k_2) + \alpha_1 \alpha_2 (k_1 + k_2)^2 + (\alpha_1 + \alpha_2)^2 k_{12}^2}}. \tag{S5}$$

Replacing $\alpha_1 \to T + \alpha + \delta\alpha$ and $\alpha_2 \to T + \alpha$ ($T$ is the temperature of the thermal bath) and Taylor-expanding for $\delta\alpha \ll 1$, we find

$$\omega_{12} = \omega_{2B} = k_{12} \frac{\sqrt{k_{12}k_2 + k_1(k_{12} + k_2)}}{k_1 + 2k_{12} + k_2} \cdot \frac{\delta\alpha}{T + \alpha} + \mathcal{O}\left(\delta\alpha^2\right). \tag{S6}$$

## TWO-BEAD MODEL PREDICTION FOR THE CYCLING FREQUENCIES: NETWORKS WITH ACTIVITY RETAINED AT TWO NODES ONLY

The two-bead model estimates of the cycling frequencies for a network with heterogeneous activity distributions are larger (in absolute value) than the simulated cycling frequencies, as mentioned in the main text and shown in Fig. 3(c). Here, we show that this deviation is due to the fluctuations of all other active beads in the network. Indeed, if we turn off the activities of all other nodes, we obtain cycling frequencies that are well predicted by the two-bead model without systematic deviations. To demonstrate this, we generate a configuration of the network where only one bead pair is "active"; all other beads are assigned zero activity, and for simplicity we set the temperature of the bath to be zero. To obtain all the cycling frequencies between different bead pairs we simulate different configurations and make a case-by-case comparison with the corresponding two-bead estimate. The results from the simulations and from the numerical solutions of the Lyapunov equation are shown in Fig. S1.

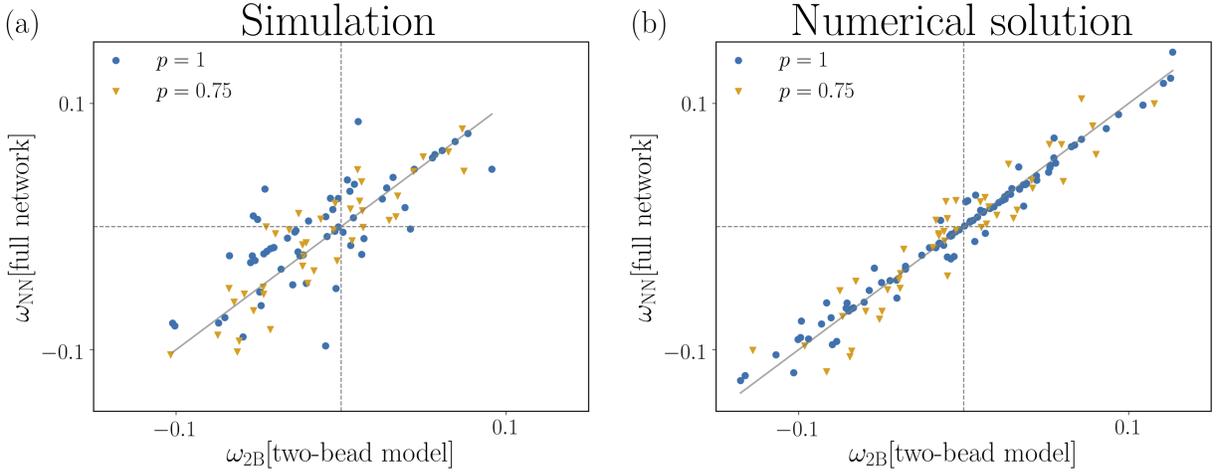

FIG. S1. (a) Scatter plot of cycling frequencies calculated from the two-bead model (x-axis) and computed via multiple simulations of a $10 \times 10$ network with one active bead pair for $p = 1$, $p = 0.75$ (y-axis). (b) The same plot as in (a) but obtained employing the numerical solution of the Lyapunov equation for the linearized system.

## DERIVATION OF EQ. (9)

The expression for the cycling frequency of our effective two-bead model follows directly from Eq. (S6) (Eq. (2) in main text) by substituting $k_{12} \to k$ and $k_1, k_2 \to \mu_m$. This yields

$$\omega_{\mathrm{EMT}} = \frac{1}{2} \frac{k}{k + \mu_m} \sqrt{\mu_m(2k + \mu_m)} \cdot \frac{\delta\alpha}{T + \alpha} + \mathcal{O}\left(\delta\alpha^2\right) . \tag{S7}$$

Note that this equation follows from a first order expansion in $\delta\alpha \ll 1$: corrections to the mean-field values of the spring constants contribute only to the second order, i.e.

$$\omega_{\mathrm{EMT}} = f(k, \mu_m) \frac{\delta\alpha}{T + \alpha} + \mathcal{O}\left(\delta k_1^2, \delta k_{12}^2, \delta k_2^2, \delta k_1 \delta\alpha, \delta k_{12} \delta\alpha, \delta k_2 \delta\alpha, \delta\alpha^2\right) . \tag{S8}$$

As noted in the main text, this explains why our mean-field predictions are in good agreement with the numerically calculated values.

## ANALYTICAL DERIVATION OF THE SCALING REGIMES FOR $\mathcal{P}_{\omega_{\mathrm{EMT}}}$

Here we describe the analytical procedure to retrieve the four different scaling regimes of $\mathcal{P}_{\omega_{\mathrm{EMT}}}$. First of all, note that the cycling frequency estimate in Eq. (9) is composed of two contributions: the stochastic contribution $\delta\alpha/(T + \alpha)$ and the elastic contribution that depends on $k$ and $\mu_m$. This elastic prefactor is completely deterministic and acts as a scale factor when computing the distribution of cycling frequencies.

For simplicity we rewrite Eq. (S7) in the following way

$$\omega_{\mathrm{EMT}} = a\left(p, k, \varepsilon\right) \cdot \frac{\delta\alpha}{T + \alpha} . \tag{S9}$$

Note $\delta\alpha \sim \mathcal{N}(0, \sqrt{2}\sigma_\alpha)$ and $T + \alpha \sim \mathcal{N}(T, \sigma_\alpha)$. We now have to compute the probability density distribution of $\omega_{\mathrm{EMT}}$: given that the stochastic part $\frac{\delta\alpha}{T+\alpha}$ of Eq. (S7) is a ratio of two Gaussian random variables, we can express the symmetric, strictly positive probability density $\varrho_Z$ of $Z := \frac{\delta\alpha}{T+\alpha}$ as

$$\varrho_\omega(\omega) = \frac{1}{a(p, k, \varepsilon)} \varrho_Z(\omega/a(p, k, \varepsilon)), \ a > 0, \tag{S10}$$

for the probability density of the cycling frequencies. Note that in general $\varrho_\omega(\omega)$ is not normal. We now have to

compute the $q$-percentile of the probability density, which is defined by

$$q = \int_{-\mathcal{P}_q}^{\mathcal{P}_q} \varrho_\omega(\omega) d\omega. \tag{S11}$$

An analytical calculation of $\mathcal{P}_q$ would require finding a closed form solution of the integral appearing in Eq. (S11) and then an analytical inversion of the resulting function. Given that our goal is not to completely solve Eq. (S11), but only to find the scaling regimes of $\mathcal{P}_q$, we can proceed as follows. By fixing the distribution of activities in our system (i.e. we fix the distribution of $z$ in Eq. (S10)), the percentiles of the $\omega$-distributions for two different sets of parameters of our system must satisfy

$$\int_{-\mathcal{P}_{q,1}/a_1}^{\mathcal{P}_{q,1}/a_1} \varrho_Z(z)dz = \int_{-\mathcal{P}_{q,1}}^{\mathcal{P}_{q,1}} \varrho_{\omega_1}(\omega_1)d\omega_1 = q = \int_{-\mathcal{P}_{q,2}}^{\mathcal{P}_{q,2}} \varrho_{\omega_2}(\omega_2)d\omega_2 = \int_{-\mathcal{P}_{q,2}/a_2}^{\mathcal{P}_{q,2}/a_2} \varrho_Z(z)dz. \tag{S12}$$

Since $\varrho_Z$ is a positive function, this equation reduces to the following requirement

$$\mathcal{P}_{q,1}/a_1 = \mathcal{P}_{q,2}/a_2. \tag{S13}$$

If we now choose realization 1 to be the fully-connected network with spring-constant $k = 1$, the elastic prefactor becomes $a_1(p = 1, k = 1, \varepsilon) = \frac{\sqrt{3}}{4}$ (at $p = 1$, $\mu_m = k$ hence $a(p = 1, k, \varepsilon) = \frac{\sqrt{3}}{4} k$; this follows immediately from Eq. (S7)). Realization 2 is instead a general $p$-dependent configuration for which the percentile reads

$$\mathcal{P}_q(p, k, \varepsilon)/\mathcal{P}_q(p = 1, k = 1) = \frac{4}{\sqrt{3}} a(p, k, \varepsilon). \tag{S14}$$

Note, this equation implies that the scaling of $\mathcal{P}_{\omega_{\text{EMT}}}$ is determined by the deterministic pre-factor $a$, since $\mathcal{P}(p = 1, k = 1)$ is a number. The elastic prefactor depends on $p$ through the effective spring constant $\mu_m$, which is determined by solving the self-consistency equation

$$\left\langle \frac{\mu_m - k}{\mu_m/p_c - \mu_m + k} \right\rangle = 0, \tag{S15}$$

where the ensemble average $\langle \cdot \rangle$ is taken over the binary distribution

$$\varrho_k(x) = p \cdot \delta(x - k) + (1 - p) \cdot \delta(x - \varepsilon). \tag{S16}$$

We thus have to solve the following equation for $\mu_m$

$$\frac{p(\mu_m - k)}{k + \mu_m/p_c - \mu_m} + \frac{(1-p)(\mu_m - \varepsilon)}{\mu_m/p_c + \varepsilon - \mu_m} = 0, \tag{S17}$$

for the different $p, k, \varepsilon$ regimes. This results in [S1]

$$\mu_m \sim \begin{cases} k & p \approx 1 \\ k\Delta p & \Delta p > 0 \\ k^{1/2}\varepsilon^{1/2} & \Delta p \approx 0 \\ |\Delta p|^{-1}\varepsilon & \Delta p < 0 \end{cases}. \tag{S18}$$

Plugging this scaling form into the definition of $a(p, k, \varepsilon)$ and employing Eq. (S14), we find that the $68_{\text{th}}$ percentile of the $\omega_{\text{EMT}}$ scales as

$$\mathcal{P}_{\omega_{\text{EMT}}}(p, k, \varepsilon) \sim \begin{cases} k & p \approx 1 \\ k|\Delta p|^{1/2} & \Delta p > 0 \\ k^{3/4}\varepsilon^{1/4} & \Delta p \approx 0 \\ k^{1/2}|\Delta p|^{-1/2}\varepsilon^{1/2} & \Delta p < 0 \end{cases}. \tag{S19}$$

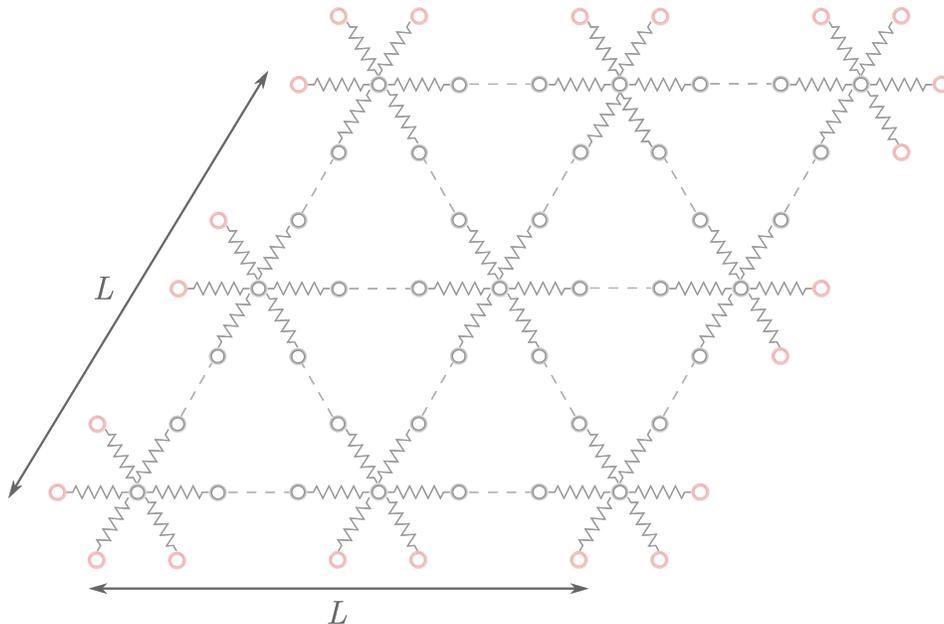

FIG. S2. Light red indicates the "frozen" nearest neighbor degrees of freedom due to the fixed boundary conditions. The system size $L$ refers to the non-fixed nodes. The different boundary scenarios of the lattice are shown: at the upper-left corner and at the lower-right corner three neighbors are fixed, while at the other two corners four neighbors are fixed. Along the four edges two neighboring nodes are absent and inside all six nearest neighbors are present.

## CALCULATION OF $p_c$ VIA COUNTING ARGUMENT

Following Maxwell's counting argument for a 2D-lattice with periodic boundary conditions [S2], we count the degrees of freedom of a triangular lattice of size $L \times L$ with fixed boundaries. This results effectively in counting the fixed neighbors of the boundary-nodes (highlighted in red in Fig. S2) and subtracting them from the result for a periodic lattice [S3].

At the four corners of the lattice there are three or four fixed neighbors depending on the corner and along the four edges each node has two fixed neighbors. In the bulk of the lattice the counting is equivalent to Maxwell's formulation. Hence, due to the corners, 14 nearest neighbors are over-counted and have to be subtracted; along the four edges there are instead $4 \cdot 2(L-2)$ neighbors that need to be subtracted. The fraction of zero-frequency modes for a 2D triangular lattice is therefore given by

$$f = \frac{2L^2 - 3pL^2 - 8p(L-2) - 14p}{2L^2} = 1 - \frac{p}{p_c},$$  (S20)

from which the critical connectivity or dilution probability is calculated to be

$$p_c = \frac{2L^2}{3L^2 + 8L - 2}.$$  (S21)

We used this to estimate $p_c$ for the finite networks presented in the main text. Note, this result is consistent with Maxwell's result $p_c = 2/3$ in the limit of an infinite lattice.

---


\end{document}